\documentclass[10pt,conference]{IEEEtran}

\usepackage{verbatim,amsmath,enumerate,amssymb,hhline}
\usepackage[square,comma,numbers]{natbib} 
\usepackage{graphicx}

\def\norm#1{\left\| #1 \right\|}
\newtheorem{definition}{Definition}[section]
\newtheorem{thm}{Theorem}[section]
\newtheorem{proposition}[thm]{Proposition}
\newtheorem{lemma}[thm]{Lemma}

\newtheorem{exam}{Example}[section]

\def\bea{\begin{IEEEeqnarray}{rCl}} 
\def\eea{\end{IEEEeqnarray}}
\def\bean{\begin{IEEEeqnarray*}{rCl}} 
\def\eean{\end{IEEEeqnarray*}}

\DeclareMathOperator*{\tr}{tr}
\DeclareMathOperator*{\nr}{nr}

\DeclareMathOperator*{\Gal}{Gal}

\providecommand{\abs}[1]{\ensuremath{\left\lvert #1 \right\rvert}}
\providecommand{\norm}[1]{\ensuremath{\left\Vert #1 \right\Vert}}

\providecommand{\vv}[1]{\textquotedblleft #1\textquotedblright}
\newcommand{\Q}{\mathbb{Q}}
\newcommand{\Z}{\mathbb{Z}}
\newcommand{\C}{\mathbb{C}}
\newcommand{\R}{\mathbb{R}}

\newcommand{\Hh}{\mathbb{H}}

\DeclareMathOperator{\SL}{SL}
\DeclareMathOperator*{\Mod}{mod}

\DeclareMathOperator{\Vol}{Vol}
\DeclareMathOperator*{\Ram}{Ram}

\providecommand{\abs}[1]{\ensuremath{\left\lvert #1 \right\rvert}}
\providecommand{\norm}[1]{\ensuremath{\left\Vert #1 \right\Vert}}

\providecommand{\vv}[1]{\textquotedblleft #1\textquotedblright}

\newtheorem{remark}{Remark}[section]

\newcommand{\D}{{\mathcal D}}

\newcommand{\A}{{\mathcal A}}

\begin{document}

\title{A new design criterion for spherically-shaped division algebra-based space-time codes}

\author{ 
\IEEEauthorblockN{Laura Luzzi}
\IEEEauthorblockA{Laboratoire ETIS,
 CNRS - ENSEA - UCP \\
Cergy-Pontoise, France \\
laura.luzzi@ensea.fr}
\and
\IEEEauthorblockN{Roope Vehkalahti}
\IEEEauthorblockA{Department of mathematics,
University of Turku\\
Finland\\
roiive@utu.fi}
}

\maketitle

\begin{abstract}
This work considers normalized inverse determinant sums as a tool for analyzing the performance of division algebra based space-time codes for multiple antenna wireless systems.
A general union bound based code design criterion is obtained as a main result.\\
In our previous work, the behavior of inverse determinant sums was analyzed using point counting techniques for Lie groups; it was shown that the asymptotic \emph{growth exponents} of these sums correctly describe the diversity-multiplexing gain trade-off of the space-time code for some multiplexing gain ranges. This paper focuses on the \emph{constant terms} of the inverse determinant sums, which capture the coding gain behavior. Pursuing the Lie group approach, a tighter asymptotic bound is derived, allowing to compute the constant terms for several classes of space-time codes appearing in the literature. \\
The resulting design criterion  suggests that the performance of division algebra based codes depends on several fundamental
algebraic invariants of the underlying algebra.
\end{abstract}


\section{Introduction}
In the last decade the problem of designing optimal space-time codes 
for the multiple-input multiple-output (MIMO) Rayleigh fading channel has attracted much attention from the coding community. Maximizing the \emph{normalized minimum determinant} of  a space-time code has been widely used as a design 
criterion.
However, this approach concentrates on minimizing the worst case pairwise error probability (PEP), and does not consider its overall distribution. The \emph{diversity-multiplexing gain trade-off} (DMT), on the other hand, 
describes
the asymptotic overall error probability as the signal-to-noise ratio and codebook size grow to infinity.  These two criteria are 
independent.
Codes with  the same DMT can have dramatically different normalized minimum determinants and vice versa. 

In \cite{VLL2013} we proposed a new criterion based on the \emph{inverse determinant sum} of the code,
which arises 
from the union bound for the PEP
\cite{TSC}. This approach forms a middle ground between DMT and normalized minimum determinant based criteria. We also proved that in many cases 
the growth of the inverse determinant sums describes the DMT of a given code for multiplexing gains $r\in [0,1]$. 

This study evidenced how the multiplicative structure of the unit group of the code comes into play; by considering the classical embedding of the unit group into a Lie group, we provided a classification of
division algebra based codes according to the growth exponent of their inverse determinant sums. 

In this paper we consider a normalized version of the inverse determinant sum, which allows us to compare the coding gains of different division algebra based codes with the same growth exponent.
 This approach takes into account both the
 number of occurrences 
 of the worst case error probability and 
 the  overall distribution. As a main result we will get a new design criterion for division algebra based space-time codes.

Our method follows the lines presented in \cite{VLL2013} combining information of the zeta-function and of the unit group of a maximal order of a division algebra. However, we tighten the previous bound and  use an explicit version of  Lie point counting from  \cite{GoNe}. A central role in the analysis is played by the \emph{Tamagawa volume formula}, which allows us to  give a detailed description of the growth of the unit group.

\section{Preliminaries}

We consider a slow fading channel with $n_t$ transmit and $n_r$  receive antennas, where the decoding delay is $T$ time units. 
The channel equation is
$Y=\sqrt{\rho/n_t}HX +N$,
where $H \in M_{n_r \times n_t}(\C)$ is the channel matrix 
and $N\in M_{n_r \times T}(\C)$
is the noise matrix. The entries of $H$ and $N$ are assumed to be 
independent identically distributed (i.i.d.) zero-mean complex circular symmetric Gaussian random variables with variance 1.
$X \in M_{n_t\times T}(\C)$ is the transmitted codeword, and $\rho$ represents the signal to noise ratio.

\subsection{Matrix Lattices and spherically shaped coding schemes}\label{latticesection}
We now suppose that $n_t=T=n$.
 
\begin{definition}
A {\em space-time lattice code} $L \subseteq M_n(\C)$ has the form
$
\Z B_1\oplus \Z B_2\oplus \cdots \oplus \Z B_k,
$
where the matrices $B_1,\dots, B_k$ are linearly independent over $\R$, \emph{i.e.}, form a lattice basis, and $k$ is
called the \emph{rank}  or the \emph{dimension} of the lattice.
\end{definition}

\begin{definition}\label{def:NVD}
If the minimum determinant of the lattice $L \subseteq M_n(\C)$ is non-zero, i.e. it satisfies
\[
\inf_{{\bf 0} \neq X \in L} \abs{\det (X)} > 0, 
\]
we say that the code has a \emph{non-vanishing determinant} (NVD).
\end{definition}

Let $\norm{\cdot}_F$ be the Frobenius norm. For 
$M>0$ we define the finite code 
$$
L(M)=\{a  \,|\,a \in L, \norm{a}_F \leq M \},
$$
and the sphere with radius $M$
$$
B(M)=\{a  \,|\,a \in M_n(\C), \norm{a}_F \leq M \}.
$$

Let $L\subseteq M_n(\C)$ be a $k$-dimensional lattice. For any fixed $m\in \Z^+$ we define 
$$
S_L^m(M):=\sum_{X\in L(M) \setminus \{\mathbf{0}\}} \frac{1}{|\det(X)|^{m}}.
$$
Our main goal is to study the growth of this sum as $M$ increases. Note, however, that in order to have a fair comparison between two different space-time codes, these should be normalized to have the same average energy. Namely, the volume $\Vol(L)$ of the fundamental parallelotope 
$$\mathcal{P}(L)=\{\alpha_1 B_1 + \alpha_2 B_2 + \ldots + \alpha_k B_k \;|\; \alpha_i \in [0,1) \;\; \forall i\}$$
should be normalized to $1$. The normalized version of the inverse determinant sums problem is then to consider the growth of the sum $\tilde{S}_L^m(M)=S_{\tilde{L}}^m(M)$ over the lattice $\tilde{L}=\Vol(L)^{-1/k}L$. Since $\tilde{L}(M)=\Vol(L)^{-1/k}L(M \Vol(L)^{1/k})$, we have 
\begin{equation}
\tilde{S}_L^m(M)= \Vol(L)^{mn/k} S^m_L(M \Vol(L)^{1/k}).
\label{normalized_sum}
\end{equation}

\subsection{Cyclic division algebras, maximal orders and zeta functions}
Let us now consider the mathematical theory that most easily gives us high dimensional NVD lattices.

Let  $E/K$ be a cyclic field extension of degree $n$ with Galois group $\Gal(E/K)= \langle\sigma\rangle $. Define a cyclic algebra  
$$
\D=(E/K,\sigma,\gamma)=E\oplus uE\oplus u^2E\oplus\cdots\oplus u^{n-1}E,
$$
where   $u\in\mathcal{D}$ is an auxiliary
generating element subject to the relations
$xu=u\sigma(x)$ for all $x\in E$ and $u^n=\gamma\in K^*$.  We assume that $\D$ is a division algebra.

Every element $x=x_0+ux_1+\cdots+u^{n-1}x_{n-1}\in\mathcal{D}$
has the following \emph{left regular representation} as a matrix $\psi(x)$:
\begin{equation*}\label{esitys}
\begin{pmatrix}
x_0& \gamma\sigma(x_{n-1})& \gamma\sigma^2(x_{n-2})&\cdots &
\gamma\sigma^{n-1}(x_1)\\
x_1&\sigma(x_0)&\gamma\sigma^2(x_{n-1})& &\gamma\sigma^{n-1}(x_2)\\
x_2& \sigma(x_1)&\sigma^2(x_0)& &\gamma\sigma^{n-1}(x_3)\\
\vdots& & & \ddots & \vdots\\
x_{n-1}& \sigma(x_{n-2})&\sigma^2(x_{n-3})&\cdots&\sigma^{n-1}(x_0)\\
\end{pmatrix}.
\end{equation*}

The  mapping $\psi$ is an injective $K$-algebra homomorphism  that allows us to identify
$\D$ with its image in $M_n(\C)$. Note that for $x \in \D$, $\det(\psi(x))=\nr(x)$, the reduced norm of $x$.

We recall here some concepts concerning the theory of orders in division algebras. Due to lack of space, we have reduced the exposition to a minimum; we refer the reader to \cite{R}.

\begin{definition}
Let $\mathcal{O}_K$ be the ring of integers of $K$. An \emph{$\mathcal{O}_K$-order} $\Lambda$ in $\D$ is a subring of $\D$, having the same identity element as $\D$, and such that $\Lambda$ is a finitely generated module over $\mathcal{O}_K$ and generates $\D$ as a linear space over $K$. 
\end{definition}

We say that $\Lambda$ is a \emph{maximal order} if it is not properly contained into any other $\mathcal{O}_K$-order of $\mathcal{D}$. \\
Let $\{w_1,\ldots,w_{n^2}\}$ be a basis of a maximal order $\Lambda$ over $\mathcal{O}_K$. The \emph{relative discriminant} of $\Lambda$ over $\mathcal{O}_K$ is defined by
$$d(\Lambda|\mathcal{O}_K)=\det\left(\tr(w_i w_j)_{i,j=1}^{n^2}\right),$$
and doesn't depend on the choice of maximal order. We denote by $\Ram_f(\mathcal{D})$ the set of primes of $\mathcal{O}_K$ which divide $d(\Lambda|\mathcal{O}_K)$, which are also called the \emph{ramified primes} \cite{R}. Moreover, for each $p \in \Ram_f(\mathcal{D})$, one can define a notion of \emph{ramification index} $1<m_p\leq n$ such that $m_p | n$ and 
\begin{equation}
d(\Lambda|\mathcal{O}_K)=\prod_{p \in \Ram_f(\mathcal{D})} p^{(m_p - 1)\frac{n^2}{m_p}}.
\label{discriminant}
\end{equation}
\smallskip \par
Given an order $\Lambda$, we define its \emph{Hey zeta function} as
\begin{equation}
\zeta_{\Lambda}(s)=\sum_{I} \frac{1}{[\Lambda:I]^s},
\label{Hey_zeta_function}
\end{equation}
where the sum is taken over all right ideals $I$ of $\Lambda$. A more explicit formula for $\zeta_{\Lambda}$ is given in \cite[p. 175]{BR2}:
\begin{equation}
\zeta_{\Lambda}(s)=\prod_{i=0}^{n-1} \zeta_K(ns-i) \prod_{p \in \Ram_f(\mathcal{D})} \prod_{\substack{0 < j \leq n-1\\ j \not\equiv 0 \Mod m_p}} (1-\mathcal{N}(p)^{j-ns}).
\label{Bushnell_Reiner_formula}
\end{equation}
Here $\zeta_K(s)$ is the \emph{Dedekind zeta function} of the center $K$, and $\mathcal{N}(p)=\abs{\mathcal{O}_K/p}$. Note that if $K=\Q(\sqrt{-d})$ is an imaginary quadratic number field, $\mathcal{N}(p)=\abs{p}^2$, and if $K=\Q$, $\mathcal{N}(p)=p$. \\
The function $\zeta_{\Lambda}(s)$ is well-defined for $\Re(s)>1$, but diverges for $s \to 1$. 
\smallskip \par
In the following we will suppose that the center $K$ of our algebra is either $\Q$ or a complex quadratic field $\Q(\sqrt{-d})$. Then $L=\psi(\Lambda)$ is a lattice in $M_n(\C)$, of dimension $k=n^2$ if $K=\Q$ and $k=2n^2$ if $K=\Q(\sqrt{-d})$, and we can consider the corresponding inverse determinant sums.

\subsection{Inverse Determinant Sums and the Unit Group}

The unit group $\Lambda^*$ of an order $\Lambda$  consists of elements $x\in \Lambda$ such that there exists an $y\in \Lambda$ with 
$xy=1_{\A}$. \\
If $K$ is $\Q$ or $\Q(\sqrt{-d})$, the units of reduced norm $1$ form a subgroup of finite index in $\Lambda^*$ \cite[p. 221]{Kl94}:
 
\begin{lemma}
The unit group $\Lambda^*$ has a subgroup
$$
\Lambda^1=\{x \,|\,x\in \Lambda^*,  \nr(x)=1\},
$$
and we have $[\Lambda^*:\Lambda^1]<\infty$.
\end{lemma}

\begin{remark} When $\mathcal{D}$ is a quaternion algebra with no real ramified places, $\nr: \Lambda^* \to \mathcal{O}_K^*$ is surjective \cite[Theorem 11.6.1]{MR} and therefore $[\Lambda^*:\Lambda^1]=\abs{\mathcal{O}_K^*}$. The cardinality $\abs{\mathcal{O}_K^*}$ is equal to $2$ if $K=\Q$ and $K=\Q(\sqrt{-d})$, except for the special cases $K=\Q(i)$ ($\abs{\mathcal{O}_K^*}$=4) and $K=\Q(e^{\frac{i\pi}{3}})$ ($\abs{\mathcal{O}_K^*}$=6). 
\label{index_Lambda1}
\end{remark}

We have shown in \cite[proof of Proposition 6.7]{VLL2013} that the growth of the inverse determinant sum for $L=\psi(\Lambda)$ is completely characterized by the growth of the unit group:
\begin{equation} 
S^{2n_r}_{\psi(\Lambda)}(M)=\sum_{x\in X(M)}\frac{|\psi(x\Lambda^*) \cap B(M)|}{|\det(\psi(x))|^{2n_r}},
\label{XM}
\end{equation}
where $X(M)$ is some collection of elements $x \in \Lambda$ such that $\norm{\psi(x)}_F\leq M$, each generating a different right ideal.\\
Let $j=[\Lambda^*:\Lambda^1]$. By choosing a set $\{a_1,\ldots,a_j\}$ of coset leaders of $\Lambda^1$ in $\Lambda^*$, we have 
\begin{equation}
S^{2n_r}_{\psi(\Lambda)}(M)= \sum_{x\in X(M)}\sum_{i=1}^j \frac{|\psi(xa_i\Lambda^1)\cap B(M)|}{|\det(\psi(x))|^{2n_r}}.
\label{bound_prop67}
\end{equation}


To obtain a good estimate of the inverse determinant sum bound, we need to study the behavior of the terms $\abs{\psi(xa_i\Lambda^1) \cap B(M)}$. This will be done in the next section using some tools from Lie group theory.

\section{Lie  Groups,  Lattices and Volumes of Spheres}
In this section we will consider a Lie group $G$, where $G$ is $\SL_n(\R)$, $\SL_n(\C)$ or $\SL_n(\Hh)$, and its \emph{arithmetic lattice subgroups}, that are
discrete  subgroups having  finite covolume. In the following we will discuss the problem of counting the number of points of these subgroups that lie inside the sphere $B(M)$. We refer the reader to \cite{GN} for the relevant definitions and an introduction to the subject.
Here we consider $\SL_n(\Hh)$ as embedded in $M_{2n}(\C)$ by replacing each quaternion element by its common $2\times 2$ matrix representation.

Each of these groups admits a multiplicative Haar measure that gives us a natural concept of volume $\Vol_G$.
In particular we can consider the volumes of the balls ${\Vol}_G (\mathcal{B}(M))$, where $\mathcal{B}(M)$ here refers to all the matrices in $G$ that have Frobenius norm smaller than $M$.

Let us now concentrate on lattice subgroups $H$ that are \emph{cocompact}, meaning that the factor group $G/H$ is compact.
In the following two results we suppose that $G$ is one of the previously mentioned Lie groups.
\begin{thm}[Corollary 1.11 and Remark 1.12, \cite{GoNe}]\label{Gorodnik}
Consider a Lie group $G$, a discrete cocompact lattice $H\subset G$ and $x\in G$. We then have that
$$
\lim_{M\to\infty}\left|\frac{ xH\cap B(M)}{\Vol_G(\mathcal{B}(M))}\right| =\frac{1}{\Vol_G(G/H)}
$$
The limit is approached uniformly for all $x\in G$.
\end{thm}
The asymptotic growth of the arithmetic lattice is thus completely determined by the volume of the ball ${\Vol}_G(\mathcal{B}(M))$. The following estimate holds:
\begin{lemma}\label{luzzi} We have that 
$${\Vol}_G(\mathcal{B}(M)) \sim C_{G} M^T,$$
where the growth exponent is 
\begin{itemize}
\item[-] $T=n^2-n$ if $G=\SL_n(\R)$,
\item[-] $T=2n^2-2n$ if $G=\SL_n(\C)$,
\item[-] $T=4n^2-4n$ if $G=\SL_n(\Hh)$.
\end{itemize}
\end{lemma}
This result is a consequence of a general theorem of 
\cite{GW}. 
The computation of these 
exponents
in the cases $\SL_n(\R)$, $\SL_n(\C)$, $\SL_n(\Hh)$ can be found in our previous work \cite[Appendix A]{VLL2013}.
\smallskip \par
We are now well-equipped to study the sum (\ref{bound_prop67}). Our reasoning follows the lines of \cite{VLL2013}, but in this paper we will obtain a tighter bound. \\
By rescaling both the discrete set and the ball, recalling that $\abs{\det(\psi(a_i))}=1$, we have 
$$\abs{\psi(xa_i\Lambda^1) \cap B(M)}=\abs{\frac{\psi(xa_i\Lambda^1)}{\det(\psi(xa_i))^{\frac{1}{n}}} \cap B\left(\frac{M}{\abs{\det(\psi(x))}^{\frac{1}{n}}}\right)}$$
Suppose that $H=\psi(\Lambda^1)$ is a cocompact lattice subgroup of $G$, where $G=\SL_n(\C),\SL_n(\R)$ or $\SL_{n/2}(\Hh)$. Note that the scaled set
$$\frac{\psi(xa_i\Lambda^1)}{\det(\psi(xa_i))^{\frac{1}{n}}}=\frac{\psi(xa_i)}{\det(\psi(xa_i))^{\frac{1}{n}}}\psi(\Lambda^1)$$
is of the form $y_iH$ with $y_i=\psi(xa_i)/\det(\psi(xa_i))^{\frac{1}{n}} \in G$.
Using Lemma \ref{luzzi}, we then have the asymptotic estimate
\begin{align}    
&\abs{\psi(xa_i\Lambda^1) \cap B(M)} \sim \frac{\Vol_G(\mathcal{B}(M \abs{\det(\psi(x))}^{-\frac{1}{n}}))}{\Vol_G(G/\psi(\Lambda^1))} \notag\\ 
& \sim \frac{C_G M^T}{\Vol_G(G/\psi(\Lambda^1)) \abs{\det(\psi(x))}^{\frac{T}{n}}}.
\label{lemma93}
\end{align}
Combining equations (\ref{bound_prop67}) and (\ref{lemma93}), we obtain
$$S_{\psi(\Lambda)}^{2n_r}(M) \sim \frac{C_G[\Lambda^*:\Lambda^1]M^T}{\Vol_G(G/\psi(\Lambda^1))} \sum_{x \in X(M)} \frac{1}{\abs{\det(\psi(x))}^{2n_r+T/n}}$$ 
Let $K=\Q$ or $\Q(\sqrt{-d})$. Since the index of a principal right ideal $x\Lambda$ of $\Lambda$ is given by
$[\Lambda:x\Lambda] = N_{\mathcal{D}/\Q}(x)=\abs{\det(\psi(x))}^{n[K:\Q]}$,
recalling the definition of the Hey zeta function (\ref{Hey_zeta_function}), we have
\begin{align*}
& \sum_{x \in X(M)} \frac{1}{\abs{\det(\psi(x))}^m} \leq \sum_{x \in X(M)} \frac{1}{[\Lambda:x\Lambda]^{\frac{m}{n[K:\Q]}}} 
\leq \\ & \leq 
\zeta_{\Lambda}\left(\frac{m}{n[K:\Q]}\right).
\end{align*}
Note that if all right ideals of $\Lambda$ are principal, then this bound is asymptotically tight. 
We can now state the following lemma: 

\begin{lemma} \label{normalized_bound}
Let $\Lambda$ be a maximal order in a division algebra $\mathcal{D}$ of degree $n$ over $K$, where $K=\Q$ or $\Q(\sqrt{-d})$, such that all right ideals of $\Lambda$ are principal. Suppose that $\psi(\Lambda^1)$ is a cocompact lattice subgroup of $G$, where $G=\SL_n(\C),\SL_n(\R)$ or $\SL_{n/2}(\Hh)$. 
Let $a=\frac{2n_r+T/n}{n[K:\Q]}>1.$ Then the normalized inverse determinant sum is asymptotically given by:
$$\tilde{S}_{\psi(\Lambda)}^{2n_r}(M) \sim \frac{C_G [\Lambda^*:\Lambda^1]\Vol(\psi(\Lambda))^{\frac{2n_rn+T}{k}}}{\Vol_G(G/\psi(\Lambda^1))} \zeta_{\Lambda}\left(a\right) M^T.$$
\end{lemma}

\section{Inverse determinant sums of central division algebras over complex quadratic fields}

Consider the case where $\mathcal{D}$ is an index $n$ $K$-central division algebra, where $K=\Q(\sqrt{-d})$ is a complex quadratic field such that $\mathcal{O}_K$ is a principal ideal domain (PID). 
The dimension of the lattice $\Lambda$ is then $k=2n^2$, and the volume of its fundamental parallelotope   is \cite{VHLR}
$$\Vol(\psi(\Lambda))=2^{-n^2} \sqrt{\abs{d(\Lambda|\Z)}}.$$

%

In the following we will denote $\SL_n(\C)$ with $G$.
Note that $\psi(\Lambda^1)\subseteq \SL_n(\C)$ and that it is a cocompact lattice subgroup \cite[Theorem 1]{Kl94}. Moreover, one can show that if $\mathcal{O}_K$ is a PID, then all right ideals of $\Lambda$ are principal \cite{R} so that Lemma \ref{normalized_bound} holds. Specializing the Lemma 
to the complex quadratic case, we obtain for $n_r>1$ 
\begin{align}
&\tilde{S}_{\psi(\Lambda)}^{2n_r}(M) \sim \frac{C_G\abs{\mathcal{O}_K^*}\abs{d(\Lambda|\Z)}^{\frac{t}{2}}}{2^{n^2t}\Vol_{G}(G/\psi(\Lambda^1))}\zeta_{\Lambda}\left(t\right) M^{2n^2-2n} 
\label{complex_quadratic_bound}
\end{align}
where $t=\frac{n_r}{n}+1-\frac{1}{n}$. Here we have used the fact that $[\Lambda^*:\Lambda^1]=\abs{\mathcal{O}_K^*}$ (Remark \ref{index_Lambda1}).

\subsection{Quaternion division algebras with complex quadratic center}

Let us now concentrate on the case where we have a quaternion division algebra ($n=2$).
Note that we have \cite{R}
$$d(\Lambda|\Z)=\abs{d(\Lambda|\mathcal{O}_K)}^2 d(\mathcal{O}_K|\Z)^{4}.$$
From the discriminant formula (\ref{discriminant}), remarking that $m_p=2$ for the ramified primes, we get 
\begin{align*} 
&\abs{d(\Lambda|\mathcal{O}_K)}=\prod_{p \in \Ram_f(\mathcal{D})} \abs{p}^2=\prod_{p \in \Ram_f(\mathcal{D})} \mathcal{N}(p)
\end{align*}

In the quaternion case, the covolume of the unit group $\Lambda^1$ in $\SL_2(\C)$ can be computed explicitly and is given by the \emph{Tamagawa volume formula} (see \cite[equation (11.2)]{MR}, and \cite[Chapitre IV, Corollaire 1.8]{Vig}):
$$\Vol_{G}(G/\psi(\Lambda^1))=\abs{d(\mathcal{O}_K|\Z)}^{\frac{3}{2}}\zeta_K(2)\prod_{p \in \Ram_f(\mathcal{D})} (\mathcal{N}(p)-1)$$ 
Let $s=\frac{n_r}{n}$, so that $t=s+\frac{1}{2}$. From equation (\ref{Bushnell_Reiner_formula}) we have
\begin{align*}
&\zeta_{\Lambda}(t)=
\zeta_K(2s+1)\zeta_K(2s) \prod_{p \in \Ram_f(\mathcal{D})} (1-\mathcal{N}(p)^{-2s}).
\end{align*}
After simplifying the expression (\ref{complex_quadratic_bound}), we obtain
\begin{align*} 
&\tilde{S}_{\psi(\Lambda)}^{2n_r}(M) \sim C_G \abs{\mathcal{O}_K^*} \abs{d(\mathcal{O}_K|\Z)}^{2s-\frac{1}{2}}  \frac{\zeta_K(2s+1)\zeta_K(2s)}{2^{4s+2}\zeta_K(2)} 
\cdot \\ &\cdot 
\prod_{p \in \Ram_f(\mathcal{D})} \frac{\mathcal{N}(p)^{s+1/2} (1-\mathcal{N}(p)^{-2s})}{\mathcal{N}(p)-1} M^4
\end{align*}
For the symmetric case $n_r=n=2$,  we finally get:
\begin{multline}
\tilde{S}_{\psi(\Lambda)}^{4}(M) \sim 
\zeta_{K}(3) C_G \abs{\mathcal{O}_K^*} 2^{-6} \abs{d(\mathcal{O}_K|\Z)}^{\frac{3}{2}} 
\cdot \\  \cdot 
\prod_{p \in \Ram_f(\mathcal{D})}  
\left(\mathcal{N}(p)^{1/2}+\mathcal{N}(p)^{-1/2}\right) 
M^4.
\label{kleinian_case}
\end{multline}
\begin{exam} Suppose $K=\Q(i)$. To find the best maximal order code according to equation (\ref{kleinian_case}), we need to minimize the product $\prod_{p \in \Ram_f(\mathcal{D})} \left(\mathcal{N}(p)^{1/2}+\mathcal{N}(p)^{-1/2}\right) $. The function $x \mapsto \sqrt{x} + 1/\sqrt{x}$ being increasing for $x \geq 1$, this can be done by choosing the 
smallest possible number of ramified primes, which is two, and the two primes with the smallest possible norm. This design criterion coincides with the one proposed in \cite{VHLR} based on the normalized minimum determinant.
\end{exam}
 
\section{Inverse determinant sums of $\Q$-central division algebras}

We now suppose that $\mathcal{D}$ is a division algebra with center $\Q$. We distinguish two main cases, depending on the ramification of the algebra at infinity. 

\begin{definition} 
Let $\D$ be an index $n$  $\Q$-central division algebra. If 
$$
\D\otimes_{\Q}\R\cong M_n(\R),
$$
we say that $\D$ is \emph{not ramified at the infinite place} (or \emph{split}).
If $2|n$ and 
$$
\D\otimes_{\Q}\R\cong M_{n/2}(\Hh),
$$
we say that  $\D$ is \emph{ramified at the infinite place}. 
\end{definition}

We will refer to the  isomorphism given  in the previous definition as $\psi_1$. The mapping $\psi_1$ has similar properties to the mapping $\psi$ obtained by the left regular representation (in particular the results about norms and lattice structure of $\psi_1(\Lambda)$ are true; see \cite{VLL2013} for more details).\\
Note that every right ideal of $\Lambda$ is principal except possibly when $\mathcal{D}$ is a quaternion algebra which is ramified at the infinite place \cite{R}. 

\subsection{Split division algebras with center $\Q$}
Suppose $K=\Q$ and $\mathcal{D} \otimes_{\Q} \R = M_n(\R)$, and let $\Lambda$ be a maximal $\Z$-order of $\mathcal{D}$. 
The dimension of the lattice $\Lambda$ is $k=n^2$, and the fundamental parallelotope has volume \cite{BCC}
$$\Vol(\psi_1(\Lambda))=\abs{d(\Lambda|\Z)}^{\frac{1}{2}}.$$ 
In the following  we will denote $\SL_n(\R)$ with $G$.
Just as before we have that $\psi_1(\Lambda^1)\subseteq G$ and that it is a cocompact lattice subgroup \cite[Theorem 1]{Kl94}. 
Specializing Lemma \ref{normalized_bound} to the split rational case, we obtain for $s=\frac{2n_r}{n}$, $t=s+1-\frac{1}{n}$,
\begin{equation*}
\tilde{S}_{\psi_1(\Lambda)}^{2n_r}(M) \sim \frac{2C_G \Vol(\psi_1(\Lambda))^{t}}{\Vol_G(G/\psi_1(\Lambda^1))}\zeta_{\Lambda}\left(t\right)M^{n^2-n} 
\end{equation*}
Here we have used the fact that $[\Lambda^*:\Lambda^1]=2$ (Remark \ref{index_Lambda1}).\\ 
If $\mathcal{D}$ is a quaternion algebra ($n=2$), we have the following Tamagawa volume formula for the unit group \cite{MR,Vig}:
$$\Vol(\SL_2(\R)/\psi_1(\Lambda^1))=\zeta(2) \prod_{p \in \Ram_f(\mathcal{D})} (p-1),$$
where $\zeta$ denotes the Riemann zeta function.
Using the formula (\ref{Bushnell_Reiner_formula}) for the Hey zeta function, we obtain 
\begin{align*}
&\tilde{S}_{\psi_1(\Lambda)}^{2n_r}(M) \sim 2C_G\frac{\zeta(2s+1) \zeta(2s)}{\zeta(2)} 
\prod_{p \in \Ram_f(\mathcal{D})} \frac{p^{s+\frac{1}{2}} (1-p^{-2s})}{p-1}M^2.
\end{align*}
When $s= 1$, corresponding to $n_r=n/2$, we get
$$\tilde{S}_{\psi_1(\Lambda)}^{n}(M) \sim2C_G \zeta(3) \prod_{p \in \Ram_f(\mathcal{D})}(p^{1/2}+p^{-1/2}) M^2 .$$

\subsection{Ramified division algebras with center $\Q$}
Suppose $K=\Q$ and $\mathcal{D} \otimes_{\Q} \R = M_{n/2}(\Hh)$, and let $\Lambda$ be a maximal $\Z$-order of $\mathcal{D}$. 
The dimension of the lattice $\Lambda$ is $k=n^2$, and its fundamental parallelotope has again volume 
$\Vol(\psi_1(\Lambda))=\abs{d(\Lambda|\Z)}^{\frac{1}{2}}$ \cite{BCC}.\\
In the following  we will denote $\SL_{n/2}(\Hh)$ with $G$. Just as before we have that $\psi_1(\Lambda^1)\subseteq G$ and that it is a cocompact lattice subgroup \cite[Theorem 1]{Kl94}. As discussed before, Lemma 3.3 holds if $n>2$. In this case, we have
for $t=\frac{2n_r}{n}+1-\frac{2}{n}$,
\begin{align*}
\tilde{S}_{\psi_1(\Lambda)}^{2n_r}(M) \sim \frac{2C_G \Vol(\psi_1(\Lambda))^{t}}{\Vol_G(G/\psi_1(\Lambda^1))}\zeta_{\Lambda}\left(t\right)M^{n^2-2n}.
\end{align*}

If $n=2$ we have growth exponent $T=0$. Indeed, the group of units $\Lambda^1$ is a finite subgroup of the compact group $\SL_1(\Hh)\cong\{a,b \in \C \;|\; \abs{a}^2 + \abs{b}^2=1\}$, which is a $4$-dimensional sphere.\\ 
Let us now suppose that $\Lambda$ has class number $1$, so that every right ideal is principal. The finite unit group changes our analysis slightly and we can use directly 
equation (\ref{XM})
to get
$$
\tilde{S}_{\psi_1(\Lambda)}^{2n_r}(M) \sim \zeta_{\Lambda}(s,\abs{d(\Lambda|\Z)}^{\frac{1}{2}}M^4) \abs{d(\Lambda|\Z)}^{\frac{n_r}{2}} \abs{\Lambda^*},
$$
where $\zeta_{\Lambda}(s,M)$ denotes the truncated Hey zeta function (over the 
ideals with index 
smaller than $M$). The bound is asymptotically tight since 
$\abs{\det(\psi_1(x))}=\norm{\psi_1(x)}_F^2/2$. \\ 
Let us now concentrate on the scenario where $n_r=1$.  The previous then transforms into
\begin{align*}\label{ramquatbound}
&\tilde{S}_{\psi_1(\Lambda)}^2(M) \sim \zeta(1,\abs{d(\Lambda|\Z)}^{\frac{1}{2}}M^4)\zeta(2) 
\cdot \notag\\ &\cdot
\prod_{p \in \Ram_f(\mathcal{D})}  
\frac{p-1}{p}
\abs{d(\Lambda|\Z)}^{\frac{1}{2}} \abs{\Lambda^*}.
\end{align*}

If $\Lambda$ has class number 1, the \emph{Eichler mass formula} gives 
$$\prod_{p \in \Ram_f(\mathcal{D})} (p-1)  \abs{\Lambda^*}= 24.$$ 
We also have that $\abs{d(\Lambda|\Z)}^{1/2}=\prod_{p \in \Ram_f(\mathcal{D})} p$.  
Equation (\ref{Bushnell_Reiner_formula}) then implies the following: 

\begin{proposition}\label{all_equal}
Let $\D$ be a $\Q$-central quaternion division algebra, which is ramified at infinity, and that $\Lambda$ is a maximal order in $\D$. If $\Lambda$ has class number $1$, then
$$
\tilde{S}_{\psi_1(\Lambda)}^2(M) \sim 24\zeta(1,\abs{d(\Lambda|\Z)}^{\frac{1}{2}}M^4)\zeta(2).
$$
\end{proposition}

Therefore, we expect all space-time codes carved from maximal orders of class number $1$ in quaternion division algebras of this type to have asymptotically the same performance when using one receive antenna. 

\begin{figure}[h]
    \centering
    \includegraphics[width=0.5\textwidth]{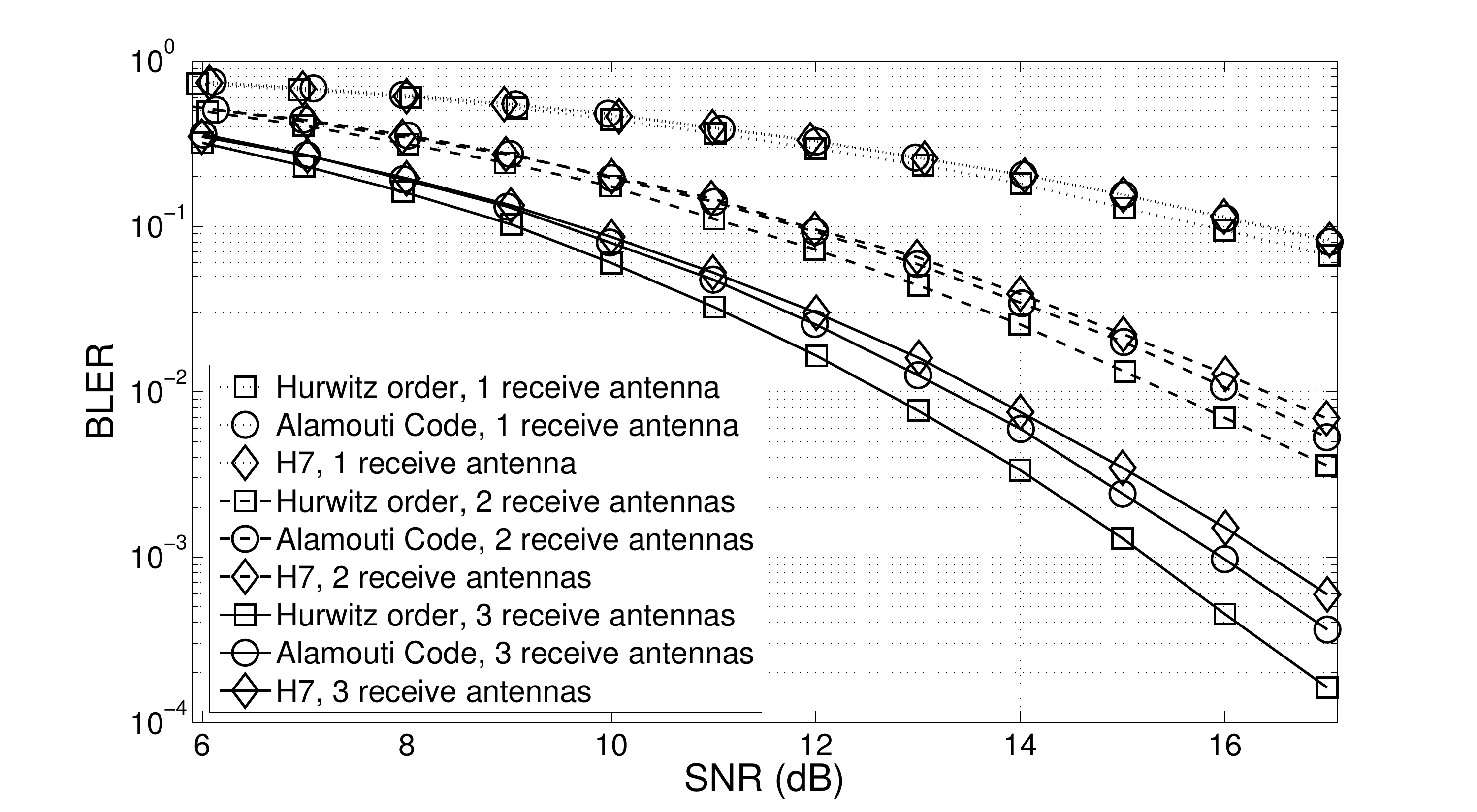}
    \caption{Simulation results for spherically-shaped codes based on quaternion algebras over $\Q$ which are ramified at infinity, using $4$-PAM constellations.}
    \label{figure}
\end{figure}

\begin{exam}
Consider the cyclic division algebras $\mathcal{H}_2=(\Q(i)/\Q,\sigma,-1)$ and $\mathcal{H}_7=(\Q(\sqrt{-7})/\Q,\sigma,-1)$, where $\sigma$ denotes complex conjugation. Note that $\Ram_f(\mathcal{H}_2)=\{2\}$, $\Ram_f(\mathcal{H}_7)=\{7\}$. The corresponding maximal orders $\Lambda_2$, $\Lambda_7$ have class number $1$. Note that the Hurwitz order $\Lambda_2$ contains the order of the Alamouti code. \\
From Proposition \ref{all_equal}, we expect similar performance for these codes for one receive antenna when using large signal constellations. For two receive antennas, we get 
$$\tilde{S}^4_{\psi_1(\Lambda)} \sim \zeta(3)\zeta(4) 24 \prod_{p \in \Ram_f(\mathcal{D})} (p+1+1/p),$$ so we expect better performance from the Hurwitz order $\Lambda_2$, which has a smaller ramified prime. Figure \ref{figure} shows that this is the case, and that the performance gap increases for $n_r=3$. Note that for finite constellations $\Lambda_2$ is still slightly better than $\Lambda_7$ even for one receive antenna.  

\end{exam}

\section*{Acknowledgements}
The research of  R. Vehkalahti is supported  by the Academy of Finland  grant \#252457. \\

\bibliographystyle{IEEEtran}

{\small
\begin{thebibliography}{10}


%

\bibitem{BCC} Eva Bayer-Fluckiger, J.-P. Cerri, J.~Chaubert, ``Euclidean minima and central division algebras,'' \emph{International Journal of Number Theory}, vol.~5, pp. 1155--1168, 2009.


\bibitem{BR2}  C. J. Bushnell, I. Reiner,  ``L-functions of arithmetic Orders and asymptotic distribution of ideals'', \emph{J. Reine Angew. Math.}, 327, pp 156--183,1981. 


\bibitem{GW} A. Gorodnik, B. Weiss, ``Distribution of lattice orbits on homogeneous varieties'', \emph{Geom.
Funct. Anal.} 17, no. 1, pp. 58--115, 2007.


\bibitem{GN} A. Gorodnik, A. Nevo, \emph{The ergodic theory of lattice subgroups}, Annals of Mathematics
Studies, vol. 172, Princeton University Press, 2010.

\bibitem{GoNe} A. Gorodnik, A. Nevo, ``Counting lattice points'', \emph{J. Reine Angew. Math.}, issue 663, pp. 127--176, 2012.


\bibitem{Kl94} E. Kleinert, \vv{Units of classical orders: a survey}, \emph{L'Enseignement Math.} 40, pp. 205--248, 1994.


\bibitem{MR} C. Maclachlan, A. W. Reid, \vv{The arithmetic of hyperbolic 3-manifolds}, Graduate texts in Mathematics, Springer, 2003
\bibitem{R}    I. Reiner, {\it Maximal Orders}, Academic Press, New York 1975.


\bibitem{TSC} V. Tarokh, N. Seshadri, A.R. Calderbank, ``Space-Time Codes for High Data Rate Wireless Communications: Performance Criterion and Code Construction'', {\it IEEE Trans.
Inf. Theory}, vol. 44, pp. 744--765, March 1998.



\bibitem{VHLR} R. Vehkalahti, C. Hollanti, J. Lahtonen, K. Ranto, ``On the densest MIMO lattices from cyclic division algebras'',
\emph{IEEE Trans.  Inf. Theory}, vol 55, no 8, pp. 3751--3780, August 2009.



\bibitem{VLL2013} R. Vehkalahti, H.-f. Lu, L. Luzzi,  ``Inverse Determinant Sums and Connections Between Fading Channel Information Theory and Algebra'', to appear in \emph{IEEE Trans. Inf. Theory}, preprint available at arXiv:1111.6289.

\bibitem{Vig} M.-F. Vign\'eras, \vv{Arithm\'etique des Alg\`ebres de Quaternions}, Lecture Notes in Mathematics, Springer Verlag 1980.

%
%
%
%
%
%
%
%
%
%

\end{thebibliography}
}
\end{document}